\documentclass[eqsecnum,showkeys,showpacs,nofootinbib,aps,epsfig]{revtex4}
\renewcommand{\theequation}{\arabic{equation}}
\usepackage[pdftex]{graphicx}
\def\beq{\begin{equation}}
\def\eeq{\end{equation}}
\def\bea{\begin{eqnarray}}
\def\eea{\end{eqnarray}}\def\nn{\nonumber}

\def\nn{\nonumber}

\begin{document}
\input epsf
\title{Free fall temperature of Schwarzschild-Tangherlini-AdS black hole}
\author{Soon-Tae Hong}
\email{soonhong@ewha.ac.kr}
\affiliation{Department of Science
Education and Research Institute for Basic Sciences, Ewha Womans
University, Seoul 120-750, Republic of Korea}
\date{\today}
\begin{abstract}
Investigating five-dimensional Schwarzschild-Tangherlini-AdS black hole, we construct 
a (6+2)-dimensional flat embedding structure. Exploiting this flat manifold, we evaluate a 
free fall temperature of the black hole in manifold ranging from outer event horizon to the infinity.  
\end{abstract}
\maketitle

%%%%%%%%%%%%%%%%%%%%%%%%%%%%%%%%%%%%%%%%%%%%%%%%%%%%%%%%%%%%%%%%%%%%%%%%
\section{Introduction}
\setcounter{equation}{0}
\renewcommand{\theequation}{\arabic{section}.\arabic{equation}}
%%%%%%%%%%%%%%%%%%%%%%%%%%%%%%%%%%%%%%%%%%%%%%%%%%%%%%%%%%%%%%%%%%%%%%%%

Since Schwarzschild-Tangherlini black hole was proposed~\cite{tan63}, considerable progress has been made 
in higher dimensional general relativity. The higher dimensional gravity was exploited to describe the black 
hole entropy for a five-dimensional black hole in string theory by counting the degeneracy of BPS
soliton bound states~\cite{vafa96}. The five-dimensional black hole 
theory admits a rotating black ring which is a stationary asymptotically flat vacuum solution associated with an 
event horizon of non-spherical topology~\cite{reall02}. Moreover, the ten-dimensional black hole possesses dynamical 
instabilities of even horizons~\cite{gregory93}. On the other hand, (2+1) dimensional Ba\~nados-Teitelboim-Zanelli 
(BTZ) black hole~\cite{btz1,btz12,btz2,cal,mann93} was proposed and it is related with the string theory in 
ten-dimensional spacetime. Moreover, a slightly modified solution of the BTZ  black hole produces a solution to 
the string theory, so-called the black string~\cite{horowitz93,horowitz932,hong01prd}. This black string solution 
is known to be a solution to the lowest order $\beta$-function equation associated with quantum corrections~\cite{callan85}.

One of the features of solutions to the field equations of general relativity is the presence of singularities. As a novel 
way to remove the coordinate singularities, higher dimensional flat embedding of the black hole solution is a subject of 
interest to both mathematicians and physicists. In differential geometry, the four-dimensional Schwarzschild 
metric~\cite{sch} is well known not to be embedded in $R^{5}$~\cite{spi75}. For the Schwarzschild black hole, a 
(5+1)-dimensional flat embedding was constructed~\cite{deser97,deser972,deser99} to remove the coordinate singularity and to 
study a thermal Hawking effect on the curved manifold~\cite{haw75,haw752} related with the Unruh effect~\cite{unruh76} in the 
higher dimensional manifold. Recently, a free fall temperature of the Schwarzschild-AdS black hole was evaluated in the higher dimensional 
flat spacetime~\cite{thorlacius08}. Later, the free fall temperature for the Gibbons-Maeda-Garfinkle-Horowitz-Strominger 
spacetime~\cite{gibbons88,gibbons882} was evaluated~\cite{park14}.

In this paper, we will construct the free fall temperature for the Schwarzschild-Tangherlini-AdS black hole. To do this, we will 
construct (6+2)-dimensional flat embedding. This paper is organized as follows: In Section II, we will explicitly evaluate the 
free fall temperature of the Schwarzschild-Tangherlini black hole, and in Section III we will derive the free fall 
temperature for the Schwarzschild-Tangherlini-AdS black hole. Section IV includes a summary and discussions.

%%%%%%%%%%%%%%%%%%%%%%%%%%%%%%%%%%%%%%%%%%%%%%%%%%%%%%%% 
\section{Free fall temperature of Schwarzschild-Tangherlini black hole}
\setcounter{equation}{0}
\renewcommand{\theequation}{\arabic{section}.\arabic{equation}}
%%%%%%%%%%%%%%%%%%%%%%%%%%%%%%%%%%%%%%%%%%%%%%%%%%%%%%%%

In order to investigate the free fall temperature of the
five-dimensional Schwarzschild-Tangherlini black hole defined on the total
manifold $S^{3}\times {\mathbf R}^{2}$, we consider a five-metric of the form \beq
ds^{2}=-N^{2}dt^{2}+N^{-2}dr^{2}+r^{2}d\Omega_{3}^{2},
\label{metric5} \eeq where the lapse function is given by~\cite{emparan08}
\beq
N^{2}=1-\frac{\mu}{r^{2}},\label{n2}
\eeq
and
\beq
d\Omega_{3}^{2}=d\alpha^{2}+\sin^{2}\alpha
(d\theta^{2}+\sin^{2}\theta d\phi^{2}). \label{n2} \eeq 
Here $\Omega_{3}$ is the solid angle in the three-dimensional compact
sphere $S^{3}$ whose value is $2\pi^{2}$. Here, we have three
angles of the three-sphere whose ranges are defined by $0\le
\alpha\le \pi$, $0\le \theta\le \pi$ and $0\le \phi\le 2\pi$. In
the lapse function $N$, we have the five-dimensional Schwarzschild-Tangherlini
radius $\mu$ defined as~\cite{emparan08} 
\beq
\mu=\frac{8GM}{3\pi}, \eeq with the black hole mass $M$.
Here we observe that in the weak gravity limit the gravitational potential in
the five-dimensional theory is proportional to $1/r^{2}$,
different from the $1/r$ of the four-dimensional Newtonian theory.

Now, we evaluate the event horizon radius $r_{H}$ by exploiting the
vanishing lapse function at $r=r_{H}$ to yield \beq
r_{H}=\mu^{1/2}, \eeq and the surface gravity $\kappa$ is given
by \beq
\kappa=\frac{1}{2}\frac{dN^{2}}{dr}|_{r=r_{H}}=\frac{1}{r_{H}}.
\eeq 
The Hawking temperature is then given by
\beq
T_{H}=\frac{1}{2\pi r_{H}}.
\label{th}
\eeq

After some algebra, using embedding isometry for the five-dimensional
Schwarzschild-Tangherlini black hole we construct the (6+1)-dimensional flat embedding structure~\cite{hongjkps14} 
\beq
ds^{2}=\eta_{AB}dz^{A}dz^{B},\label{ds2}
\eeq
where the (6+1) flat metric is given by
\beq
\eta_{AB}={\rm diag}~(-1,+1,+1,+1,+1,+1,+1)
\label{gemsds} 
\eeq 
with the coordinate transformations \bea
z^{0}&=&\kappa^{-1}\left(1-\frac{\mu}{r^{2}}\right)^{1/2}\sinh \kappa t,\nn\\
z^{1}&=&\kappa^{-1}\left(1-\frac{\mu}{r^{2}}\right)^{1/2}\cosh \kappa t,\nn\\
z^{2}&=&\int dr~\left[\frac{r_{H}^{2}(r^{3}+r_{H}r^{2}+r_{H}^{2}r+r_{H}^{3})}{r^{4}(r+r_{H})}
\right]^{1/2},\nn\\
z^{3}&=&r\sin\alpha\sin\theta\cos\phi,\nn\\
z^{4}&=&r\sin\alpha\sin\theta\sin\phi,\nn\\
z^{5}&=&r\sin\alpha\cos\theta,\nn\\
z^{6}&=&r\cos\alpha. \label{gems} 
\eea

Now we assume that an observer at rest is freely falling from the radial position $r=r_{0}$ 
at $\tau=0$~\cite{thorlacius08,park14}. The equations of motion for the orbit of the observer are given as follows
\bea
\frac{dt}{d\tau}&=&\frac{(1-x_{0})^{1/2}}{1-x},\nn\\
\frac{dr}{d\tau}&=&-(x-x_{0})^{1/2},\label{eomr}
\eea 
where the dimensionless variable $x$ is defined as
\beq
x=\frac{\mu}{r^{2}}.
\label{xdimless}
\eeq
Exploiting Eqs. (\ref{gems}) and (\ref{eomr}), we obtain $a_{7}$ acceleration:
\beq
a_{7}^{2}=\frac{1}{\mu}(1+x+x^{2}).\label{a7}
\eeq
Using the definition of the freely falling observer at rest (FFAR) temperature
\beq
T_{FFAR}=\frac{a_{7}}{2\pi},
\eeq
we end up with
\beq
T_{FFAR}^{2}=\frac{1}{4\pi^{2}\mu}(1+x+x^{2}).
\label{tffar0}
\eeq

%%%%%%%%%%%%%%%%%%%%%%%%%%%%%%%%%%%%%%%%%%%%%%%%%%%%%%%%%%%%%%%%%%%%%%%%%%%%%
\begin{figure}[!t]
\begin{center}
\epsfysize=4.5cm
\includegraphics[width=7cm]{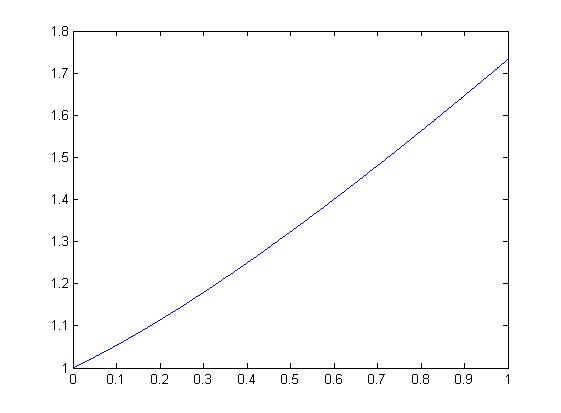}\\
%\epsfbox{fig1eps.eps}\\
\end{center}
\vskip -0.5cm 
\caption[fig1] {$T_{FFAR}/T_{H}$ in terms of the dimensionless variable
$x=\mu/r^{2}$.} \label{fig1}
\end{figure}
%%%%%%%%%%%%%%%%%%%%%%%%%%%%%%%%%%%%%%%%%%%%%%%%%%%%%%%%%%%%%%%%%%%%%%%%%%%%%

In Figure 1, we depict the ratio of the free fall temperature $T_{FFAR}$ and the Hawking 
one $T_{H}$ as a function of the dimensionless variable $x$ defined in Eq. (\ref{xdimless}). 
Here one notes that, at $x=0$ corresponding to the asymptotic region, the free fall 
temperature approaches to the Hawking temperature, as expected. At the event horizon $r=\mu^{1/2}$,
we find the temperature $T_{FFAR}=\sqrt{3}T_{H}$. This value is different from that in the 
four-dimensional Schwarzschild black hole case where the free fall temperature is given by 
$T_{FFAR}^{4d}=2T_{H}^{4d}$~\cite{thorlacius08}. Next, we obtain the local fiducial temperature 
corresponding to the Unruh temperature for the Schwarzschild-Tangherlini black hole:
\beq
T_{FID}^{2}=\frac{1}{4\pi^{2}\mu(1-x)}.
\label{tfid}
\eeq

%%%%%%%%%%%%%%%%%%%%%%%%%%%%%%%%%%%%%%%%%%%%%%%%%%%%%%%%%%%%%%%%%%%%%%%%
\section{Free fall temperature of Schwarzschild-Tangherlini-AdS black hole}
\setcounter{equation}{0}
\renewcommand{\theequation}{\arabic{section}.\arabic{equation}}
%%%%%%%%%%%%%%%%%%%%%%%%%%%%%%%%%%%%%%%%%%%%%%%%%%%%%%%%%%%%%%%%%%%%%%%%

In this section, we will derive the free fall temperature for the 
Schwarzschild-Tangherlini-AdS black hole. To do this, we first construct the 
higher dimensional flat embedding of this black hole. We next exploit the orbit equations for the black hole
to obtain the corresponding free fall temperature. We start with the Schwarzschild-Tangherlini-AdS 
black hole described by the lapse function~\cite{maeda14}
\beq
N^{2}=1-\frac{\mu}{r^{2}}+\frac{r^{2}}{l^{2}},\label{n2ads}
\eeq
where $\Lambda=l^{-2}$ is the cosmological constant. The lapse function can be also written in 
terms of the outer horizon $r_{+}$ as follows
\beq
N^{2}=\frac{(r^{2}-r_{+}^{2})(r^{2}+r_{+}^{2}+l^{2})}{l^{2}r^{2}}.
\label{lapsesta}
\eeq 
Adopting the dimensionless variables 
\beq
x=\frac{r_{+}^{2}}{r^{2}},~~~c=\frac{l}{r_{+}},
\eeq
we rewrite the lapse function (\ref{lapsesta}) as follows
\beq
N^{2}=\frac{(1-x)[1+(c^{2}+1)x]}{c^{2}x}.
\eeq

After some algebraic manipulation, we newly find the (6+2)-dimensional flat embedding with the metric of the form 
\beq
\eta_{AB}={\rm diag}~(-1,+1,+1,+1,+1,+1,+1,-1)
\eeq
associated with the coordinate transformations \bea
z^{0}&=&\kappa^{-1}\left(1-\frac{\mu}{r^{2}}+\frac{r^{2}}{l^{2}}\right)^{1/2}\sinh \kappa t,\nn\\
z^{1}&=&\kappa^{-1}\left(1-\frac{\mu}{r^{2}}\frac{r^{2}}{l^{2}}\right)^{1/2}\cosh \kappa t,\nn\\
z^{2}&=&\int dr~\left[\frac{l^{2}r_{+}^{2}(r^{3}+r_{+}r^{2}+r_{+}^{2}r+r_{+}^{3})}{r^{4}(r+r_{+})
(r^{2}+r_{+}^{2}+l^{2})}
\right]^{1/2},\nn\\
z^{3}&=&r\sin\alpha\sin\theta\cos\phi,\nn\\
z^{4}&=&r\sin\alpha\sin\theta\sin\phi,\nn\\
z^{5}&=&r\sin\alpha\cos\theta,\nn\\
z^{6}&=&r\cos\alpha,\nn\\
z^{7}&=&\int dr~\left[\frac{\kappa^{-2}(r^{4}+3r_{+}^{4}+2l^{2}r_{+}^{2})
(r^{3}+r_{+}r^{2}+r_{+}^{2}r+r_{+}^{3})+l^{2}r^{4}(r+r_{+})(r^{2}+r_{+}^{2})}{l^{2}r^{4}(r+r_{+})
(r^{2}+r_{+}^{2}+l^{2})}
\right]^{1/2},\label{gemsads}
\eea
where the surface gravity for the Schwarzschild-Tangherlini-AdS black hole is given by
\beq
\kappa=\frac{2r_{+}^{2}+l^{2}}{l^{2}r_{+}}
\eeq
Moreover, the Hawking temperature is given by
\beq
T_{H}=\frac{2r_{+}^{2}+l^{2}}{2\pi l^{2}r_{+}}.
\label{thads}
\eeq

%%%%%%%%%%%%%%%%%%%%%%%%%%%%%%%%%%%%%%%%%%%%%%%%%%%%%%%%%%%%%%%%%%%%%%%%%%%%%
\begin{figure}[!t]
\begin{center}
\epsfysize=4.5cm
\includegraphics[width=7cm]{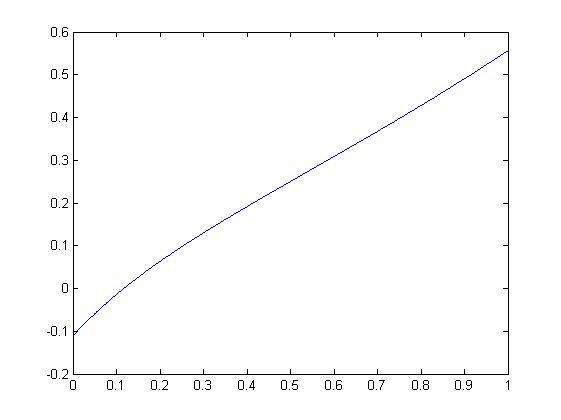}\\
%\epsfbox{fig1eps.eps}\\
\end{center}
\vskip -0.5cm 
\caption[fig1] {$T_{FFAR}^{2}/T_{H}^{2}$ for $c=1$ in terms of the dimensionless variable
$x=\mu/r^{2}$.} \label{fig1}
\end{figure}
%%%%%%%%%%%%%%%%%%%%%%%%%%%%%%%%%%%%%%%%%%%%%%%%%%%%%%%%%%%%%%%%%%%%%%%%%%%%%

Following the algorithm for finding the free fall temperature discussed above, we construct the equations of the 
orbit for the free fall observer at rest at $r=r_{0}$
\bea
\frac{dt}{d\tau}&=&\frac{cx(1-x_{0})^{1/2}[1+(c^{2}+1)x_{0}]^{1/2}}{x_{0}^{1/2}(1-x)[1+(c^{2}+1)x]},\nn\\
\frac{dr}{d\tau}&=&-\left[\frac{(1-x_{0})[1+(c^{2}+1)x_{0}]}{c^{2}x_{0}}-\frac{(1-x)[1+(c^{2}+1)x]}{c^{2}x}\right]^{1/2}.
\label{orbitads}
\eea
Exploiting the coordinate transformations (\ref{gemsads}) and the orbit equations (\ref{orbitads}), 
after some tedius algebra we arrive at the free fall temperature for the Schwarzschild-Tangherlini-AdS black hole: 
\beq
T_{FFAR}^{2}=\frac{-1+(c^{2}+1)(c^{2}+3)x+(c^{2}+1)^{2}(1+x)x^{2}}{4\pi^{2}l^{2}[1+(c^{2}+1)x]}.
\label{tffar}
\eeq 
Here we note that in the vanising $c$ limit the free fall temperature is reducible to that of the Schwarzschild-Tangherlini black hole 
in Eq. (\ref{tffar0}). Moreover, in the very large $c$ limit, the free fall temperature becomes 
\beq
\frac{T_{FFAR}^{2}}{T_{H}^{2}}=1+x+x^{2},
\eeq
where $T_{H}$ is the Hawking temperature given in Eq. (\ref{thads}). Figure 2 shows the ratio of $T_{FFAR}^{2}$ 
and $T_{H}^{2}$ for the case of $c=1$ as a function of the dimensionless variable $x$. Finally, we evaluate 
the local fiducial temperature corresponding to the Unruh temperature for the Schwarzschild-Tangherlini-AdS black hole as 
follows
\beq
T_{FID}^{2}=\frac{(2+c^{2})^{2}x}{4\pi^{2}l^{2}(1-x)[1+(1+c^{2})x]}.
\label{tfidads}
\eeq

%%%%%%%%%%%%%%%%%%%%%%%%%%%%%%%%%%%%%%%%%%%%%%%%%%%%%%%%
\section{Conclusions}
\setcounter{equation}{0}
\renewcommand{\theequation}{\arabic{section}.\arabic{equation}}
%%%%%%%%%%%%%%%%%%%%%%%%%%%%%%%%%%%%%%%%%%%%%%%%%%%%%%%%

In summary, we have constructed the (6+2)-dimensional flat embedding space for the 
five-dimensional Schwarzschild-Tangherlini-AdS black hole. Using this flat embedding manifold, we have 
calculated the free fall temperature of the black hole in manifold ranging from outer event horizon to the infinity.  

Now we have couple of comments to address. In the coordinate transformations (\ref{gems}) and (\ref{gemsads}), 
there exist no coordinate singularities at event horizons $r=r_{H}$ and $r=r_{+}$, respectively. The free fall 
temperatures in Eqs. (\ref{tffar0}) and (\ref{tffar}) are 
well defined without any divergence features. However, the local fiducial temperature in Eq. (\ref{tfid}) for the 
Schwarzschild-Tangherlini balck hole diverges at the event horizons $r=r_{H}$ corresponding 
to $x=1$. As $r$ approaches infinity, the local fiducial temperature becomes the Hawking temperature. 
For the Schwarzschild-Tangherlini-AdS balck hole case, the local fiducial temperature in Eq. (\ref{tfidads}) vanishes in 
asymptotically flat spacetime and also diverges at the event horizons $r=r_{+}$.

\end{document}